\PassOptionsToPackage{cmyk,table}{xcolor}
\documentclass[conference,flushend]{iaria}

\usepackage[babel=true,english=british]{csquotes}
\usepackage[british]{babel}

\usepackage[shortcuts]{extdash}

\usepackage[alwaysadjust]{paralist}
\usepackage[caption=false,font=footnotesize]{subfig}

\usepackage[
babel=true,
expansion=alltext,
protrusion=alltext-nott,
nopatch=eqnum,
final
]{microtype}

\usepackage{threeparttable}
\usepackage{array}
\newcolumntype{C}[1]{>{\centering\arraybackslash\hspace{0pt}}p{#1}}
\usepackage{amsmath,amssymb,amsfonts}
\usepackage{algorithmic}
\usepackage{graphicx}
\usepackage{textcomp}
\usepackage{xcolor}
\usepackage{booktabs}
\usepackage{balance}
\usepackage{fontawesome}
\usepackage{relsize}
\usepackage{makecell}
\usepackage{multirow}
\usepackage{tikz}
\tikzset{>=latex}
\usetikzlibrary{arrows,positioning,shapes.geometric,calc}
\usepackage{mathabx}
\usepackage{comment}
\usepackage{balance}

\usepackage[acronym,nonumberlist,nopostdot]{glossaries}
\newacronym{CAPEC}{CAPEC}{Common Attack Pattern Enumeration and Classification}
\newacronym{CPDFD}{CPDFD}{Cyber-Physical Data Flow Diagram}
\newacronym{CPS}{CPS}{Cyber-Physical System}
\newacronym{CVE}{CVE}{Common Vulnerabilities and Exposures}
\newacronym{CWE}{CWE}{Common Weakness Enumeration}
\newacronym{DFD}{DFD}{Data Flow Diagram}
\newacronym{DoS}{DoS}{Denial of Service}
\newacronym{ICS}{ICS}{Industrial Control System}
\newacronym{ID}{ID}{Identifier}
\newacronym{IoT}{IoT}{Internet of Things}
\newacronym{IP}{IP}{Intellectual Property}
\newacronym{GSM}{GSM}{Global System for Mobile Communications}
\newacronym{HID}{HID}{Human Interface Device}
\newacronym{HWD}{HWD}{Hardware Diagram}
\newacronym{NA}{N/A}{Not Applicable}
\newacronym{OTP}{OTP}{One Time Programmable}
\newacronym{OWASP}{OWASP}{Open Web Application Security Project}
\newacronym{PCB}{PCB}{Printed Circuit Board}
\newacronym{RFID}{RFID}{Radio-Frequency Identification}
\newacronym{ROM}{ROM}{Read-Only Memory}
\newacronym{SoC}{SoC}{System on a Chip}
\newacronym{SRAM}{SRAM}{Static Random Access Memory}
\newacronym{SWD}{SWD}{Serial Wire Debug}
\newacronym{TEE}{TEE}{Trusted Execution Environment}
\newacronym{TPM}{TPM}{Trusted Platform Module}
\newacronym{TTM}{TTM}{Thing Threat Modelling}

\let\sub\textsubscript

\newcommand{\GCPDFD}{G\textsubscript{CPDFD}}
\newcommand{\GDFD}{G\textsubscript{DFD}}

\newcommand{\dCohen}[1]{$d\textsubscript{\textit{Cohen}} = #1$}
\newcommand{\MWUT}{Mann-Whitney \textit{U} test}
\newcommand{\SWT}{Shapiro-Wilk test}
\newcommand{\tT}{\textit{t}-test}

\title{Introducing the Cyber-Physical Data Flow Diagram to Improve Threat Modelling of Internet of Things Devices}

\author{%
    \large 
    Simon Liebl\,$^{1,2}$\,\orcidlink{0000-0003-1311-4202}, Ian Ferguson\,$^2$\,\orcidlink{0000-0001-9866-6182}, Andreas Aßmuth\,$^3$\,\orcidlink{0009-0002-2081-2455}, Natalie Coull\,$^2$\,\orcidlink{0000-0003-0681-9888}, George R. S. Weir\,$^4$\,\orcidlink{0000-0002-6264-4480}\\[0.3ex]\normalsize\normalfont
    $^1$\,emgarde, Ebermannsdorf, Germany\\
    e-mail: {\tt simon.liebl@emgarde.de}\\
    $^2$\,Abertay University, Dundee, UK\\
    e-mail: {$\lbrace$\tt i.ferguson\,|\,n.coull$\rbrace$\tt @abertay.ac.uk}\\
    $^3$\,Kiel University of Applied Sciences, Kiel, Germany\\
    e-mail: {\tt andreas.assmuth@haw-kiel.de}\\
    $^4$\,University of Strathclyde, Glasgow, UK\\
    e-mail: {\tt george.weir@strath.ac.uk}
}

\usepackage{ifpdf}
\ifpdf
\pdfoutput=1 % we are running pdflatex 
\pdftrue
\fi

\makeatletter
\def\ps@IEEEtitlepagestyle{
	\def\@oddfoot{\mycopyrightnotice}
	\def\@evenfoot{}
}
\def\mycopyrightnotice{
	{\footnotesize
		\begin{minipage}{0.8\textwidth}
			\centering
			Please cite as: Simon Liebl, Ian Ferguson, Andreas Aßmuth, Natalie Coull, and George R.~S. Weir, ``Introducing the Cyber-Physical Data Flow Diagram to Improve Threat Modelling of Internet of Things Devices,'' in \emph{Proc of the First International Conference on Cross-Domain Security in Distributed, Intelligent and Critical Systems (CROSS-SEC 2026), Lisbon, Portugal, pp.~30--39, April 2026.}
		\end{minipage}
	}
}
\makeatother
\begin{document}

\maketitle

\begin{abstract}
    A growing number of Internet of Things (IoT) devices are used across consumer, medical, and industrial domains. They interact with their environment through sensors and actuators and connect to networks such as the Internet. Because sensors may collect sensitive data and actuators can trigger physical actions, security, privacy, and safety are major challenges. Threat modelling can help identify risks, but established IT‑focused methods transfer to the IoT only to a limited extent. In this paper, a new modelling technique specifically for IoT devices called Cyber-Physical Data Flow Diagram (CPDFD) is proposed that also allows modelling of hardware with the aim to support manufacturers in identifying threats and developing countermeasures. The technique was examined through an experimental study and a survey with interviews. The results suggest that numerous other attack scenarios can be found through the modelling technique, improving the identification of threats to IoT devices.
\end{abstract}


\section{Introduction}
The \gls{IoT} is the interconnection of billions of devices via the Internet or another network. These devices are used in a wide variety of areas, such as the smart home, medical, and industrial applications. Their key characteristics include interacting with the environment via sensors and actuators and communicating with other devices and systems. Networking them aims to extend functionality, integrate services, improve usability, increase efficiency, and reduce costs. However, adding multiple network interfaces also expands the attack surface and increases the risk of cyber attacks. It seems that the number of published incidents is steadily increasing \cite{IoTincreasingAtt}, indicating that \gls{IoT} security is a serious problem. In recent years, incidents have been reported in various \gls{IoT} application fields, such as industrial (e.g., attacks on Ukraine's power grid \cite{UkraineGrid}), medical (e.g., insulin pumps and ventilators \cite{BSIMedIoT}), automotive (e.g., hack of Tesla and its Wall Connector \cite{Tesla3rdParty,TeslaWallConnector}), infrastructure (e.g., traffic lights \cite{GreenLights}), and smart home (e.g., vacuum robot \cite{giese_reverse_2024}).

The described small sample of security incidents related to \gls{IoT} devices and systems shows that there is a systematic problem threatening security, privacy, and safety. Therefore, new approaches are required to better protect \gls{IoT} devices from such threats. A promising approach that is already established in the IT domain is the risk assessment using threat modelling. Threat modelling is the systematic approach to identifying threats and vulnerabilities to a particular system. Appropriate countermeasures can then be defined. The process can also be carried out early in the product lifecycle, which supports the goal security by design. However, common threat modelling approaches from the IT domain cannot necessarily be adopted to \gls{IoT} devices. Reasons for this are that \gls{IoT} devices consist of resource-constrained embedded systems and that their interaction with the real world poses an increased risk to the privacy and safety of users. As part of this work, the modelling technique \gls{DFD}, which is often used in threat modelling, was specifically examined. Several adaptations are proposed to increase the applicability of \glspl{DFD} to \gls{IoT} devices, including two new elements and the possibility to create a \gls{HWD}. This extension, called \gls{CPDFD}, has the aim to enable more detailed modelling of the components of \gls{IoT} devices and thus allow better threat identification compared to regular \glspl{DFD}. This technique is intended to support device manufacturers in identifying threats and developing appropriate countermeasures.

The remainder of this paper is structured as follows: in \autoref{sec:Background}, the fundamentals of \gls{IoT} devices and threat modelling are provided. \autoref{sec:RelatedWork} presents related work regarding \gls{IoT} security, threat modelling, and \glspl{DFD}. In \autoref{sec:CPDFD}, the proposed extension \gls{CPDFD} is introduced. The methodology for evaluating \glspl{CPDFD} consisting of two studies is described in \autoref{sec:Meth}, followed by the presentation of the results. These results are then discussed in \autoref{sec:Discussion}. The paper ends with conclusions in \autoref{sec:Conclusions}.

\section{Background}
\label{sec:Background}
\subsection{Internet of Things Devices}
The \gls{IoT} is a ``group of infrastructures interconnecting connecting objects and allowing their management, data mining and the access to the data they generate'' \cite{IoTDefTax}. These objects are devices extended with network connectivity and computing capabilities and require usually minimal human intervention \cite{IoTOverview}. The architecture of the \gls{IoT} is often structured into different layers \cite{EnablingIoT}. The perception layer, the lowest layer, consists of these physical objects, such as sensors, actuators, or \gls{RFID} tags. These device are also used in a \gls{CPS}, for example, to control physical processes in the real world \cite{IIoTAnalysis}. For such systems, real-time requirements are essential, which differentiates them clearly from IT systems. The impact of \gls{IoT} devices can thus affect security, privacy, and safety and have consequences such as the disclosure of sensitive data in medical applications or human injury in industrial applications.

\gls{IoT} devices are embedded systems, which are specific computer systems built for a custom purpose. Their basic components consist of hardware, software, and data \cite{Liebl_Journal}. Examples for hardware components are microcontroller, memory chip, \gls{PCB}, security chip, power supply, and various sensors and actuators. Software components include firmware, basic libraries for cryptography and logging, and protocol stacks for numerous \gls{IoT} protocols, such as Bluetooth, ZigBee, and MQTT. Besides firmware, devices store access data, keys, and configuration and log files, among others.

\subsection{Threat Modelling}
In this section, the \gls{DFD} and two published threat modelling techniques are introduced. The small excerpt of techniques is necessary to understand the remainder of this paper.

\subsubsection{Data Flow Diagram}
The \gls{DFD} is a graphical modelling technique used historically in the field of software engineering and system analysis \cite{DeMarco}. It quickly gained popularity due to its intuitive nature and ability to capture both high-level and detailed views of system processes and data. A \gls{DFD} consists only of four elements: \textit{Process}, \textit{Data Store}, \textit{External Entity}, and \textit{Data Flow}. This enables a simple visual representation of complex systems and illustrates how data is input, processed, stored, and output. Processes represent the transformation of data and can be thought of any running code. Data stores depict any type of data storage, such as files or databases. External entities are used to describe external data sources or destinations, e.g., people or any code outside your control. Data flows are used to connect the other three elements. Besides their use in software engineering, security experts found their application also valuable in the field of cyber security. The analysis of the system is an important part of a security assessment conducted to find threats and vulnerabilities. In order to limit the assessment for complex systems, another element called \textit{Trust Boundary} or \textit{Trust Area} was added to visualise different trust levels, initiating the second version of \glspl{DFD} \cite{Shostack-2014}. Since \glspl{DFD} are typically created during early design, they naturally support the analysis of architectural security issues rather than implementation‑level defects.

\subsubsection{STRIDE and LINDDUN}
The CIA triad -- Confidentiality, Integrity, and Availability -- forms the basis of information security, often complemented by goals such as authenticity, non‑repudiation, and authorisation. STRIDE covers the threats spoofing, tampering, repudiation, information disclosure, denial of service, and elevation of privilege \cite{Shostack-2014}, and was developed at Microsoft to identify such threats during design \cite{STRIDE}. It maps directly to these six security goals.

STRIDE can be applied to \glspl{DFD}, but checking all six threats for every element is time‑consuming. Since some elements are more susceptible to specific threats, STRIDE‑per‑Element and STRIDE‑per‑Interaction were introduced to streamline the analysis \cite{Shostack-2014}.

LINDDUN is a framework for privacy threat modelling. Similar to STRIDE, LINDDUN is an acronym that stands for linkability, identifiability, non-repudiation, detectability, disclosure of information, unawareness, and non-compliance and can be combined with \glspl{DFD} as well \cite{LINDDUN-orig}. The threats are therefore not focused on security goals, but are instead aimed at privacy goals.

\section{Related Work}
\label{sec:RelatedWork}
% IoT security, threat modelling, DFDs
\gls{IoT} security and privacy is still a huge problem, which is why a lot of research is being conducted in a variety of directions. Due to the many articles around \gls{IoT} security, there are several reviews with the aim to summarise, among others, threats, attacks, challenges, and solutions \cite{Harbi2019,Alaba2017,Leloglu2017,Abiodun2021,Scott2023}. Some researchers focused specifically on \gls{IoT} devices, often using hands-on approaches to show how embedded systems can be attacked, which threats arise, and how they can be mitigated \cite{Arias2015,Kolias2016,Wurm2016,Atamli2014,Iskhakov2018}. Regarding threat modelling and risk assessment, there are a couple of articles with the aim of adapting these methods for the \gls{IoT}. In \cite{TMAutomationIoT}, they aim to automate this process for \gls{IoT} systems by providing their own methodology. The authors of \cite{CPSSTRIDE} present a STRIDE and \gls{DFD}-based threat modelling approach specifically for \glspl{CPS} with special focus on human injury, equipment damage as well as black-out. Many articles address specific fields of application such as automotive \cite{TowardsTMVehicles}, building and home automation \cite{TMBuilding}, agriculture \cite{TMAgriculture}, and healthcare devices \cite{TMHealthDev}. \glspl{DFD} are also relevant to this work. In this regard, \cite{berger_automatically_2016} and \cite{SolutionAwareDFDs} are particularly worth mentioning, as they propose adjustments to \glspl{DFD} with the aim of improving threat identification and generation.

In summary, there is a lot of research in the area of \gls{IoT} security and privacy, also with special emphasis on threat modelling. In the case of \gls{IoT} devices in particular, the differences between embedded systems and IT systems were recognised and threats analysed that emerge through sensors, actuators, and interaction with the environment and other systems. However, the hands-on approaches and the various methods for the different \gls{IoT} applications emphasise that a common technique for modelling \gls{IoT} devices has not yet been established. 

\section{Cyber-Physical Data Flow Diagram}
\label{sec:CPDFD}
This section introduces the proposed modelling technique \gls{CPDFD} by describing its aims and adaptations.

\subsection{Aim}
As mentioned before, there are several issues with modelling \gls{IoT} devices. One big difference between IT and \gls{IoT} applications is the interaction with the physical world. While this is an integral part of the latter, it is hardly present in IT applications. In \gls{IoT} systems, a sensor, the data source, usually measures an environmental value, such as temperature, photographic picture, or heart activity. The processing of the data can in turn lead to physical operations by the actuator, the data sink, and thus result in changes in the environment. For example, a valve is opened, a car battery is charged, or a door is unlocked. From a security point of view, these sensor values and actuator actions are particularly interesting, as they can threaten the privacy and safety of users. It is therefore necessary to address this difference, the interaction with the physical environment, also in the modelling.

Another difference between IT and \gls{IoT} applications is the location and purpose. \gls{IoT} devices are used in critical applications, such as healthcare, automotive, and automation. Many of them are placed outdoors, e.g., security cameras, charging stations, or pipeline valves. The criticality as well as the accessibility of the devices make hardware attacks interesting for attackers and are frequently reported \cite{BSI-HW-Attacks}. Successful attacks give deep insight into the device and often lead to access to \gls{IP}, sensitive data, credentials, and cryptographic secrets. Therefore, hardware attacks, which are often out of scope in IT applications, need to be considered due to the accessibility and the frequent critical application.

One core element of \gls{IoT} devices is the communication with other devices and systems. Besides dozens of wireless communication protocols, such as Z-Wave, Wi-Fi, and NFC, wired protocols including Modbus, Ethernet, and USB are commonly used. Additionally, several common protocols used in embedded systems are used, for instance, UART, RS-232, and JTAG. With the multitude of protocols and interfaces in use, it is therefore easy to lose track which of them are actually utilised in an \gls{IoT} device. However, these various protocols and interfaces increase the attack surface and can become a gateway for attackers. The goal is therefore to be able to represent these interfaces in a model in order to get a quick overview of the connectivity of a device.

There is another difference between the development of \gls{IoT} devices and IT applications. Modelling the latter is mainly done by software and network engineers that commonly have a computer science background. However, modelling an \gls{IoT} device may also require the help of hardware and embedded engineers with a background frequently in mechanical or electrical engineering. Therefore, a modelling technique for \gls{IoT} devices needs to be simple and at the same time comprehensive in order to be applicable by people with different backgrounds. 

\subsection{The Technique}

\begin{table*}[tb]
    \centering
    \def \off{\addlinespace[1mm]}
    \setlength{\tabcolsep}{5pt}
    \caption{The elements of a CPDFD. The last two columns indicate their use in DFDs and HWDs.}
    \label{tab:SOL-CPDFD-Elements}
    \footnotesize
    \vspace*{-1mm}
    \begin{tabular}{p{20mm}C{20mm}lcc}
        \toprule
        Element & Symbol & Meaning \& Examples & DFD & HWD\\ \midrule
        Process & 
        \makecell[c]{\begin{tikzpicture}
                \draw[rounded corners] (0, 0) rectangle (1.5, 0.7);
                \node[text width=1.5cm, align=center] at ({1.5/2},0.35) {Name};
        \end{tikzpicture}} 
        & \makecell[l]{Executed code\\Log. Process: Code in C, web server\\Phy. Processor: Microprocessor, TPM} & \makecell[c]{\\$\boxtimes$\\$\boxvoid$} & \makecell[c]{\\$\boxvoid$\\$\boxtimes$} \\ \off
        Data Store &  
        \makecell[c]{\begin{tikzpicture}
                \node[draw=black,cylinder,aspect=0.2,shape border rotate=90,minimum width=1.5cm] at (0,0) {Name};
        \end{tikzpicture}}
        & \makecell[l]{Things that store data\\Log. Data Store: File, database\\Phy. Data Store: FLASH, OTP memory} & \makecell[c]{\\$\boxtimes$\\$\boxvoid$} & \makecell[c]{\\$\boxvoid$\\$\boxtimes$} \\ \off
        \makecell[l]{External Entity\\Interactor} & 
        \makecell[c]{\begin{tikzpicture}
                \draw[] (0, 0) rectangle (1.5, 0.7);
                \node[text width=1.5cm, align=center] at ({1.5/2},0.35) {Name};
        \end{tikzpicture}}
        & \makecell[l]{People or code outside your control\\Log. External Entity: Browser, cloud service\\Phy. External Entity: User, machine} & \makecell[c]{\\$\boxtimes$\\$\boxtimes$} & \makecell[c]{\\$\boxvoid$\\$\boxvoid$} \\ \off
        Data Flow &  
        \makecell[c]{\begin{tikzpicture}
                \draw (0.15,0) -- (1.75,0);
                \draw (0,0) -- (0.15,-0.1) -- (0.15,0.1) -- cycle;
                \draw[fill=black] (1.75,0) -- (1.6,-0.1) -- (1.6,0.1) -- cycle;
                \node[text width=1.75cm, align=center] at ({1.75/2},0.2) {Name};
                \draw (0,0.5) -- (1.75,0.5);
                \draw[fill=black] (1.75,0.5) -- (1.6,0.4) -- (1.6,0.6) -- cycle;
                \node[text width=1.75cm, align=center] at ({1.75/2},0.7) {Name};
        \end{tikzpicture}}
        & \makecell[l]{Communication between elements\\(Log.) Data Flow: Data flow with protocol (stack)} & \makecell[c]{\\$\boxtimes$} & \makecell[c]{\\$\boxvoid$} \\ \off
        Trust Area & 
        \makecell[c]{\begin{tikzpicture}
                \draw[dashed] (0, 0) rectangle (1.5, 0.7);
                \node[text width=1.5cm, align=left] at ({1.6/2},0.5) {Name};
        \end{tikzpicture}}
        & \makecell[l]{Delimited area of trust\\Log. Trust Area: Net segment, container\\Phy. Trust Area: PCB, casing} & \makecell[c]{\\$\boxtimes$\\$\boxtimes$} & \makecell[c]{\\$\boxvoid$\\$\boxtimes$} \\ \off
        Physical Link & 
        \makecell[c]{\begin{tikzpicture}
                \draw (0,0) -- (1.35,0) -- (1.5,0.7) -- (0.15,0.7) -- cycle;
                \node[text width=1.5cm, align=center] at ({1.5/2},0.35) {Name};
        \end{tikzpicture}}
        & \makecell[l]{Link between physical and logical world\\Physical Link: Microphone, battery, motor} & \makecell[c]{\\$\boxtimes$} & \makecell[c]{\\$\boxtimes$} \\ \off
        Interface & 
        \makecell[c]{\begin{tikzpicture}
                \draw (0,0) -- (1.35,0) -- (1.5,0.35) -- (1.35,0.7) -- (0,0.7) -- (0.15,0.35) -- cycle;
                \node[text width=1.5cm, align=center] at ({1.5/2},{0.7/2}) {Name};
        \end{tikzpicture}}
        & \makecell[l]{Communication interface\\Interface: Wi-Fi, USB, JTAG} & \makecell[c]{\\$\boxvoid$} & \makecell[c]{\\$\boxtimes$} \\
        \bottomrule
    \end{tabular}
\end{table*}

As mentioned before, the \gls{CPDFD} is proposed to support modelling of \gls{IoT} devices. The technique extends the \gls{DFD}, which is commonly used for security assessments and is known by many security experts and engineers. The following paragraphs describe adjustments to the \gls{DFD} in order to achieve the previously described aims.

It is difficult to represent the interaction with the physical environment through sensors and actuators using the five elements of a \gls{DFD}. Depending on the sensor characteristics (e.g., simple electrical resistors or modules with a serial interface), it can make sense to represent them as \textit{Process}, \textit{Data Store}, or \textit{External Entity} or even as a combination with \textit{Data Flows}. However, this question of detail would take a lot of time and contradict the principle of simplicity of \glspl{DFD}. If one does not know how to represent a specific component right away, it could be omitted. This in turn would lead to a situation where measured values which might threaten the privacy, for example, would not be recorded. This applies analogously to actuators and safety as a protection goal. It is therefore proposed to add a new element called \textit{Physical Link} to the \gls{DFD} notation in order to be able to model these interactions with the physical environment. The goal of this element is to give sensors and actuators a first-class representation in the diagrams, highlighting issues with the goals of privacy and safety.

Another aim of the \gls{CPDFD} technique is to enable the creation of hardware models in order to support the identification of hardware attacks, among others. In theory, there are dozens of modelling techniques (see \cite{ESDesign}). However, the authors are not aware of any hardware modelling technique that is commonly used for the threat analysis. Hardware components were integrated into \glspl{DFD} in a few cases (e.g., \cite{PlaybookMed,Wolf2019}). It is proposed to use the \gls{DFD} notation to create a separate \gls{HWD}. This is achieved by splitting the elements into logical and physical ones. Regular \glspl{DFD} are mainly used for modelling software-related components, which are now referred to as logical elements. The additional physical elements are used for the creation of the \gls{HWD}. The physical data store can be used to model a flash or \gls{OTP} memory, for instance. Likewise, the physical processor and the physical trust area can be used to model microprocessors and \glspl{PCB}, for example. Modelling of data flows is not necessary in the \gls{HWD}.

The \gls{CPDFD} technique should also allow users to get a quick overview of the device's communication interfaces. Similar to the \textit{Physical Link} before, modelling the interfaces through one of the standard elements could have the effect that these are not modelled. It is therefore proposed to model them as a separate element named \textit{Interface} as part of the \gls{HWD}. 

All elements of a \gls{CPDFD} and their use in a \gls{DFD} and \gls{HWD} are shown in \autoref{tab:SOL-CPDFD-Elements}. The proposed extension has further advantages. A \gls{DFD} can be augmented with links to the hardware model. For example, for data stores, such as a file, it can be specified in which memory it is stored and for data flows, the utilised communication interface can be provided. This additional context can lead to a more precise description of an attack scenarios and thus support the threat identification. The new elements \textit{Physical Link} and \textit{Interface} have another benefit when it comes to software-aided modelling. Threat rules for automatic threat generation can be specified more precisely or can be created specifically for the two elements. This can reduce false positives, i.e., wrongly identified threats, and false negatives, i.e., undetected threats. The commonly used techniques STRIDE-per-Element and LINDDUN can still be used for \glspl{CPDFD} and also provide more accurate results. Last but not least, the \gls{CPDFD} technique establishes a unifying notation for \glspl{HWD} and \glspl{DFD}. The resulting diagrams remain simple and clear, despite the addition of two elements and the distinction between physical and logical elements. They can be created and understood by hardware, embedded, and software engineers and other people as well.

\section{Methodology}
\label{sec:Meth}
The methodology for evaluating the improvements of the \gls{CPDFD} is introduced in this section. It consists of an experimental study and a survey with interviews.

\subsection{Experimental Study}
\label{ssec:ExpStudy-Meth}
This section introduces an experimental study with the objective to compare \glspl{CPDFD} with \glspl{DFD} by quantitatively comparing the number of attack scenarios identified by two groups, each using one modelling technique. Participants of the study, who were students of computer science-related study programmes, were put in the following situation: As innovative students, they have applied their previously acquired knowledge in practice and developed a smart \gls{IoT} device alongside their studies. They now want to bring the device to market, but they still have concerns about security and privacy. Therefore, each participant had the task of examining the device for threats and vulnerabilities. The analysed device was the open source device Jaimico \cite{Jaimico}, which is a combination of voice assistant and health monitor. The companion robot can be worn on the shoulders in daily life and can be used as a voice assistant or health monitor to measure body temperature and heart activity by connecting to a wearable. In addition to analysing health data in the cloud, the device offers the special feature of detecting Covid19 through analysing recorded coughing and body temperature. In this study, participants were supported by a custom software tool called TTModeler \cite{TTModelerEmgarde, TTModelerGitHub}, but the supported modelling technique, specifically the two new elements, differed for both groups. The modelling technique thus serves as an independent variable. The produced project file and a concluding questionnaire served as the data source for further analysis. For the sake of completeness, it should be noted that there was a third group with a different software tool, but this is not of interest for this paper.

The aim of this study is to gather evidence on the usability of each tool, which is the ``degree to which a product or system can be used by specified users to achieve specified goals with effectiveness, efficiency, and satisfaction in a specified context of use'' \cite{ISO25022}. For this purpose, effectiveness is defined as ``accuracy and completeness with which users achieve specified goals'' and efficiency as the resources expended in relation to the effectiveness \cite{ISO25022}. More precisely, it was measured how much time is spent on the creation of a specific model and how many attack scenarios were found in this context, enabling the quantification in identified attack scenarios per minute for the hardware model, among others. This can be used to determine the efficiency. The impact of the additional elements \textit{Physical Link} and \textit{Interface} was specifically examined. 

It was decided to conduct the study with students because they are available at short notice and in larger numbers than, for example, developers from industry. Students from study programmes related to computer science were invited to participate in the study. All participating students were in the fourth year of their studies. They represent the actual users well, as they have, for example, only rudimentary experience in the field of security, as it is the case with many firmware and software developers. A total of $46$ students volunteered to participate in the study. Five of them took part in a preliminary study, which was conducted to find issues in the procedure and task description, and their results are not considered. This results in a sample size of $41$. Due to the omitted third group and unexpected cancellations, $12$ participants used the modelling technique \gls{CPDFD}, denoted as \GCPDFD, and $14$ participants used the \gls{DFD} technique (\GDFD).

Participants first read an information document covering the CIA, STRIDE, and LINDDUN threat models, as well as IoT privacy threats defined by \cite{Ziegeldorf-2014}. They then received a second document introducing \gls{DFD}s, including STRIDE‑per‑Element, LINDDUN‑per‑Element, and example diagrams. Group \GCPDFD{} received a modified version explaining \gls{CPDFD} elements instead of standard \gls{DFD}s. Next, participants read the description of the example device Jaimico and watched a video demonstrating the main features of the assigned tool, including how to model the ``Register device'' use case and add threats. They then worked on the task, with the option to stop at any time within a three‑hour limit. Finally, they completed the questionnaire and submitted their project file.

The submitted project files and questionnaires served as the data sources. Project files were analysed for diagrams, elements, and attack scenarios. Duplicates and out‑of‑scope scenarios were marked and excluded. Remaining scenarios were filtered and labelled as considered attack scenarios. Because the device description was partly fictitious, scenarios were classified as having either low or high probability of correctness. Most were rated high, with exceptions such as ``Malicious file on USB flash drive'', since the USB port was specified as charging‑only. Each scenario was assigned a type: \textit{generic}, \textit{outlined}, or \textit{custom}, indicating the level of detail and difficulty of identification. \textit{Generic} scenarios were directly derived from STRIDE or LINDDUN, \textit{outlined} scenarios added some detail (e.g., ``Clear text data storage of the database''), and \textit{custom} scenarios were highly specific or not obviously derived from a threat model (e.g., ``Unsecure Bluetooth version''). Only high‑probability scenarios of type \textit{outlined} or \textit{custom} were considered relevant.

The tool was modified to track modelling time. After completing the task, participants answered a 21‑item questionnaire, including eight items on threat identification and 13 on usability.

The further analysis of the collected data was carried out with the help of descriptive and inferential statistical methods \cite{DataAnalysisPlan} and may be visualised using a box plot \cite{BoxPlot}. The significance level for rejecting a null hypothesis is set at $\alpha = 5\%$. Depending on whether the results are normally distributed, either the \tT{} or the \MWUT{} is used to test for a difference between two groups. Subsequently, Cohen's $d$ is calculated to quantify the effect size \cite{Cohen, Hartung}. It can be assumed that there is a small effect for $0.2 < d < 0.5$, a medium effect for $0.5 \leq d < 0.8$, and a large effect for $d \geq 0.8$.

\subsection{Survey and Interviews}
\label{ssec:Survey-Meth}
The last study aims to get feedback from people who might actually utilise \glspl{CPDFD}. These include device manufacturers as well as consultancies. Manufacturers often do not have security experts themselves and therefore hire external companies. Participants apply the methodology (\gls{TTM}), technique (\gls{CPDFD}), and tool (TTModeler) on their own devices and provide afterwards feedback through a questionnaire and an interview. The aim is to specifically ask for opinions on the \gls{HWD} and the new elements.

All persons with a connection to cyber security, \gls{IoT}, or embedded systems were invited for the study. In total, $15$ participants could be recruited for the study. \autoref{tab:Survey-Participants} categorizes participants according to the type of company, the size of the company (according to \cite{DefSME}), and the experience. Noteworthy is that two participants had no cyber security experience ($13.3\%$).

\begin{table}[tb]
    \centering
    \caption{Overview of participant and company characteristics.}
    \label{tab:Survey-Participants}
    \footnotesize
    \setlength{\tabcolsep}{3pt}
    \vspace*{-1mm}
    \begin{tabular}{ll}
        \toprule
        Category & Distribution \\
        \midrule
        Company type & $8$ device manufacturers; $7$ service providers \\
        Participant role & $9$ security experts; $6$ engineers \\
        Company size & $7$ large; $3$ medium-sized; $4$ small; $1$ micro \\
        Field of application & \makecell[l]{$8$ industrial; $3$ consumer; $2$ automotive;\\$1$ medical; $1$ infrastructure} \\
        Experience & \makecell[l]{$46.7\%$ embedded security; $80.0\%$ DFD;\\$60.0\%$ threat modelling}\\
        \bottomrule
    \end{tabular}
\end{table}

At the start of the study, participants received an introduction to threat modelling, the adaptations used in this work, and their task. After completing and evaluating the contributions, they filled out a questionnaire and answered a few interview questions. The questionnaire, consisting mainly of five‑point Likert items \cite{Likert} and free‑text fields, together with the interview, served as the data sources. Questionnaire responses were analysed descriptively, while the free‑text fields and interview data were examined qualitatively.

\section{Results}
\label{sec:Results}
The following section summarises the results of both studies.

\subsection{Experimental Study}
\label{ssec:ExperimentalStudy-Res}
This section presents the results of the experimental study, summarised in \autoref{tab:TPS-Detailed-Results}. The first tests on the results using the \SWT{} showed that the sample is not normally distributed. This could be caused by the rather small sample size, for example. It was therefore decided to consistently use non-parametric methods.

\begin{table}[tb]
    \centering
    \begin{threeparttable}
        \footnotesize
        \setlength{\tabcolsep}{3pt}
        \caption{Overview of the project file analysis.}
        \label{tab:TPS-Detailed-Results}
        \vspace*{-1mm}
        \begin{tabular}{lcc}
            \toprule
            & \GCPDFD{} & \GDFD{} \\\midrule
Duration & $111$ min $\pm$ $20$ min & $120$ min $\pm$ $10$ min \\
Considered attack scenarios & $69.21 \pm 40.44$ & $45.17 \pm 19.85$ \\\hline
\multirow{2}{*}{Type: Generic}& $19.29 \pm 22.08$ & $17.92 \pm 14.64$ \\
& $27.86\% \pm 31.90\%$ & $39.67\% \pm 32.41\%$ \\
\multirow{2}{*}{Type: Outlined}& $10.86 \pm 5.84$ & $9.58 \pm 7.95$ \\
& $15.69\% \pm 8.44\%$ & $21.22\% \pm 17.60\%$ \\
\multirow{2}{*}{Type: Custom}& $39.07 \pm 17.17$ & $17.67 \pm 10.00$ \\
& $56.45\% \pm 24.80\%$ & $39.11\% \pm 22.15\%$ \\
\multirow{2}{*}{Relevant attack scenarios}& $49.93 \pm 21.00$ & $27.25 \pm 11.32$ \\
& $72.14\% \pm 30.34\%$ & $60.33\% \pm 25.07\%$ \\\hline
\multirow{2}{*}{Goal: Security}& $44.93 \pm 19.48$ & $27.00 \pm 11.14$ \\
& $89.99\% \pm 39.01\%$ & $99.08\% \pm 40.89\%$ \\
\multirow{2}{*}{Goal: Privacy}& $5.00 \pm 2.11$ & $0.25 \pm 0.45$ \\
& $10.01\% \pm 4.23\%$ & $0.92\% \pm 1.66\%$ \\
\multirow{2}{*}{\textit{Physical Link}-related}& $10.86 \pm 5.17$ &  \\
& $21.75\% \pm 10.36\%$ & \\
\multirow{2}{*}{\textit{Interface}-related}& $11.21 \pm 5.10$ &  \\
& $22.46\% \pm 10.22\%$ & \\
\multirow{2}{*}{Model: Hardware}& $29.07 \pm 11.28$ & $9.50 \pm 6.67$ \\
& $58.23\% \pm 22.58\%$ & $34.86\% \pm 24.47\%$ \\
\multirow{2}{*}{Model: Software}& $4.29 \pm 8.09$ & $4.67 \pm 5.77$ \\
& $8.58\% \pm 16.20\%$ & $17.13\% \pm 21.19\%$ \\
\multirow{2}{*}{Model: Data flow}& $16.57 \pm 8.71$ & $13.08 \pm 9.40$ \\
& $33.19\% \pm 17.44\%$ & $48.01\% \pm 34.51\%$ \\
Time for hardware model & $26$ min $\pm$ $14$ min & $12$ min $\pm$ $12$ min \\
Time for software model & $9$ min $\pm$ $7$ min & $8$ min $\pm$ $7$ min \\
Time for data flow model & $64$ min $\pm$ $21$ min & $96$ min $\pm$ $17$ min \\
Time per attack scenario & $2$ min $\pm$ $0$ min & $4$ min $\pm$ $0$ min \\
\quad - for hardware model & $1$ min $\pm$ $0$ min & $1$ min $\pm$ $1$ min \\
\quad - for software model & $2$ min $\pm$ $2$ min & $2$ min $\pm$ $2$ min \\
\quad - for data flow model & $4$ min $\pm$ $1$ min & $7$ min $\pm$ $1$ min \\
            \bottomrule
        \end{tabular}
        \begin{tablenotes}
            \footnotesize
            \item \parbox[t]{\dimexpr\linewidth}{Note: The values of the last two columns show the mean value and the standard deviation. Furthermore, the table is split in three sections. The percentages in the second section refer to the considered attack scenarios, those in the third section to the relevant attack scenarios.} 
        \end{tablenotes}
    \end{threeparttable}
\end{table}

\autoref{tab:TPS-Detailed-Results} shows the duration needed to conduct the study for both groups. Note that this duration represents the pure interaction time with the tool and does not include the learning time required for the threat models, \gls{CPDFD}/\gls{DFD}, and the introduction to Jaimico. Group \GCPDFD{} has a wide range of $80$ to $142$ minutes ($M = 111$~min, $SD = 20$ min). In contrast, \GDFD{} has a higher mean value of $120$ minutes ($SD = 10$ min) and a narrower range of $98$ to $132$ minutes. Despite these differences, there is a large overlap of the ranges and it indicates that the technique used does not have a significant influence on the execution duration (\MWUT, $U = 62.5$, $p = .280$, $p > \alpha$).

The number of considered attack scenarios, i.e., without duplicates, high probability of correctness and out of scope (see \autoref{ssec:ExpStudy-Meth}), have a large range. Participants of \GCPDFD{} found in total between $2$ and $182$ attack scenarios ($M = 69.21$, $SD = 40.44$) and \GDFD{} found between $2$ and $74$ scenarios ($M = 45.17$, $SD = 19.85$). The scenarios start at $2$, as some participants had mainly modelled the example use case showed in the introductory video. Their type, \textit{generic}, \textit{outlined}, and \textit{custom}, can be further analysed. The results for the types \textit{generic} and \textit{outlined} are similar and with overlapping error bars (see \autoref{tab:TPS-Detailed-Results}). A difference can be seen for the type \textit{custom}. The groups \GCPDFD{} and \GDFD{} found on average $39.07$ ($SD = 17.17$) and $17.67$ ($SD = 10.00$) scenarios respectively. The \MWUT{} shows that both groups are significantly different from each other. Therefore, \GCPDFD{} found significantly more attack scenarios of the high ranked type \textit{custom} than \GDFD{} ($U = 155.0$, $p = .000$, $p \leq \alpha$, \dCohen{1.50}, large effect). This means that \GCPDFD{} found more scenarios with high information value that are more difficult to find.

\begin{figure}[tb]
    \centering
    \includegraphics[width=\linewidth, trim=0cm 0.4cm 0cm 0cm, clip]{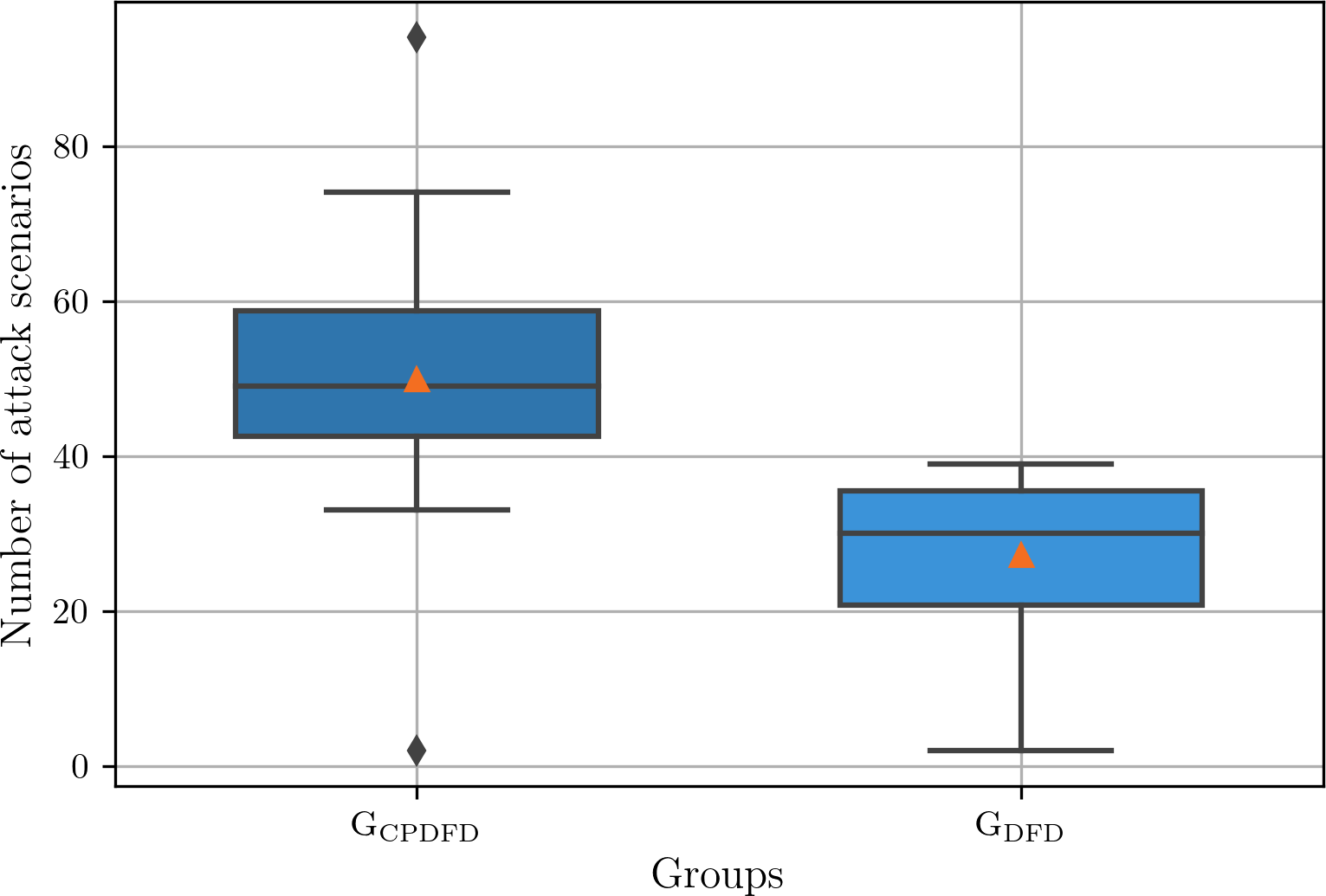}
    \vspace{-5mm}
    \caption{Relevant attack scenarios for groups \GCPDFD{} and \GDFD{} as box plot.}
    \label{fig:TPS-RelevantAS}
\end{figure}

The attack scenarios were filtered again to include only those of the type \textit{outlined} and \textit{custom} as explained in \autoref{ssec:ExpStudy-Meth}. This reduced list of relevant attack scenarios is visualised in \autoref{fig:TPS-RelevantAS}. The main range of \GCPDFD{} is between $33$ and $74$ scenarios with two outliers at $2$ and $94$, while \GDFD{} ranges between $2$ and $39$. The ranges overlap, but the boxes highlighting the lower and upper quartiles do not. The median of \GCPDFD{} ($49$) differs from \GDFD{} ($30$) by $19$ scenarios. The result of the \MWUT{} shows that group \GCPDFD{} identified significantly more attack scenarios than \GDFD{} ($U = 147.0$, $p = .001$, $p \leq \alpha$, \dCohen{1.32}, large effect). This clearly shows that group \GCPDFD, which had the \textit{Physical Link} and \textit{Interface} available, identified substantially more attack scenarios.

Attack scenarios were categorised whether these threaten the security or privacy of the device under consideration. Group \GCPDFD{} identified $5.00$ scenarios on average ($SD = 2.11$), while \GDFD{} identified less than $1$ scenario on average ($M = 0.25$, $SD = 0.45$). Therefore, they identified significantly more privacy attack scenarios than \GDFD{} (\MWUT, $U = 160.5$, $p = .000$, $p \leq \alpha$, \dCohen{3.01}, large effect). A more detailed analysis shows that out of these $5$ scenarios of \GCPDFD, $3.64$ scenarios were identified through the element \textit{Physical Link} ($SD = 1.69$), $1.21$ through the \textit{Interface} ($SD = 0.58$), and only $0.14$ scenarios through other elements ($SD = 0.36$). Consequently, this suggests that the used technique, \gls{CPDFD} or \gls{DFD}, has an influence on the identified attack scenarios against privacy.

\begin{figure}[tb]
    \centering
    \includegraphics[width=\linewidth, trim=0cm 0.4cm 0cm 0cm, clip]{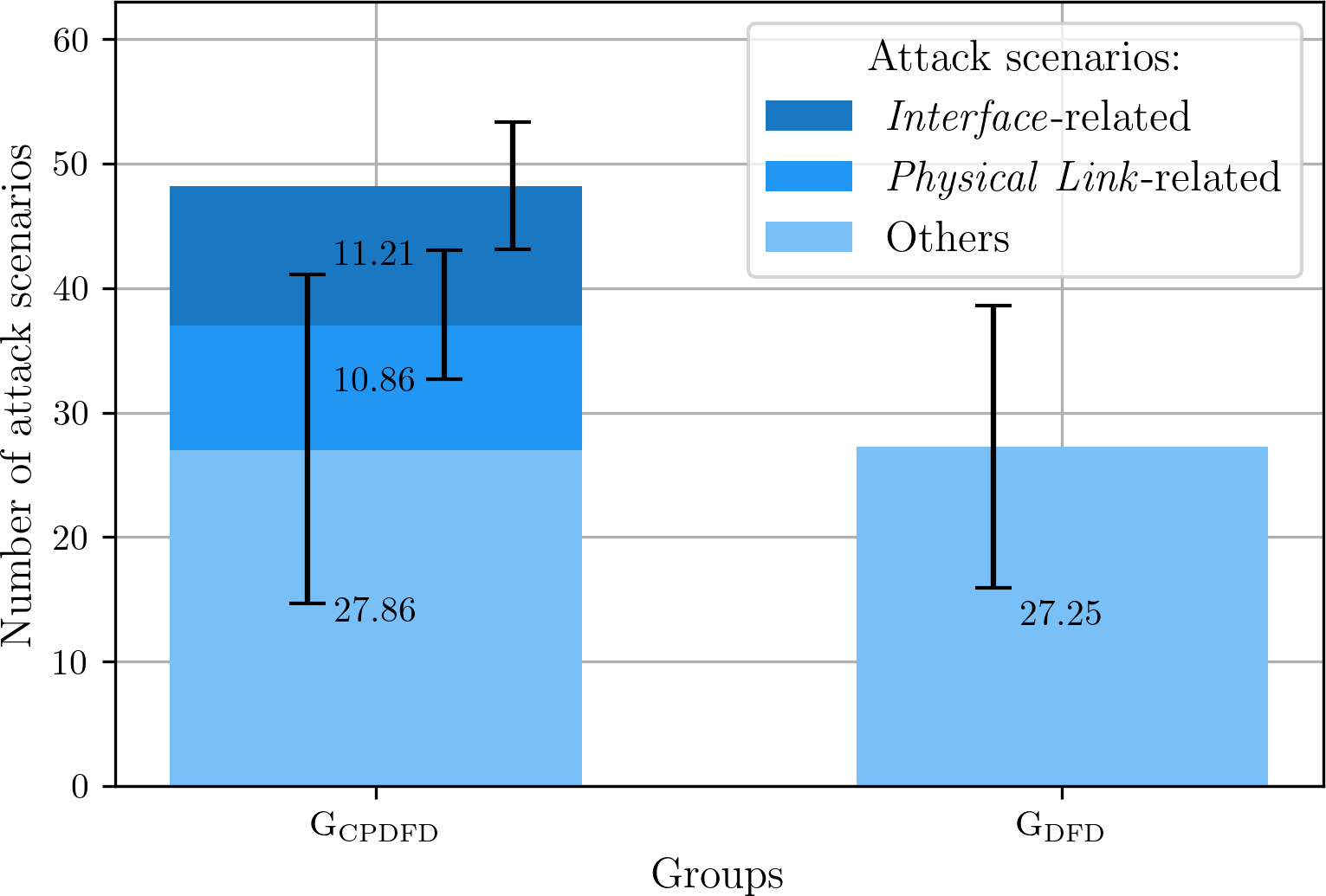}
    \vspace{-5mm}
    \caption{Average attack scenarios per group. In group \GCPDFD, scenarios relating to the elements \textit{Physical Link} and \textit{Interface} are highlighted.}
    \label{fig:TPS-PhysicalLinkInterfaceRelated}
\end{figure}

\autoref{fig:TPS-PhysicalLinkInterfaceRelated} shows the average number of identified attack scenarios per group. Furthermore, it highlights how many scenarios of \GCPDFD{} relate to the additional elements of a \gls{CPDFD} -- \textit{Physical Link} and \textit{Interface}. The comparison of \GCPDFD{} and \GDFD{} shows that both groups found on average almost the same number of scenarios for category \textit{Others}. The extra scenarios of \GCPDFD{} relate to the new elements. Therefore, this suggests that further attack scenarios can be found through the two elements.

\autoref{tab:TPS-Detailed-Results} includes in which model, hardware, software, or data flow, the attack scenarios were identified. The main difference can be seen for the hardware model. Group \GCPDFD{} has a range of $14$ to $45$ identified scenarios, without a single outlier at $2$, and a median of $29.5$. \GDFD{} has a range of $0$ to $19$ and a median of $11.5$. The \MWUT{} shows that the groups are significantly different ($U = 156.0$, $p = .000$, $p \leq \alpha$, \dCohen{2.08}, large effect). Again, the difference between \GCPDFD{} and \GDFD{} arises from the two new elements. The median \textit{Physical Link} and \textit{Interface}-related scenarios for the former are $6$ and $10.5$ scenarios, respectively. The median scenarios for the standard elements are $13$ and similar to \GDFD{}. The results indicate that it is worth creating a hardware model and that there is an improvement due to the new elements. 

Furthermore, the tool TTModeler tracked how much time participants spent on creating and analysing the different models, enabling the calculation of the duration per attack scenario (see \autoref{tab:TPS-Detailed-Results}). Group \GCPDFD{} needed on average $2$ minutes per scenario ($SD = 0$ min), while \GDFD{} needed $4$ minutes ($SD = 0$ min). The average time per attack scenario in the hardware model was even less for both groups, because both needed $1$ minute per scenario on average. These scores indicate the improved efficiency of \glspl{CPDFD} and also of a hardware model in general.

It was recorded how participants modelled certain components. The technique \gls{CPDFD} would classify microphone, loudspeaker, touch screen, and temperature sensor as element \textit{Physical Link}, for example. While almost all participants in group \GCPDFD{} modelled the microphone as \textit{Physical Link} ($93\%$), the other group was divided. One quarter modelled the microphone as \textit{Process}, $17\%$ as \textit{External Entity}, and the majority of $58\%$ did not model it at all. For the loudspeaker, the picture is almost the same: only one participant had modelled it as \textit{Data Store} instead of \textit{Process}. While the majority of \GCPDFD{} also modelled the touch screen, the temperature sensor, and the \gls{PCB}, the share for \GDFD{} was one third or less. The battery was not modelled by anyone of \GDFD, but by $86\%$ of \GCPDFD. Likewise, the same was analysed for the interfaces JTAG, Bluetooth, Wi-Fi, and USB. Almost all participants of \GCPDFD{} modelled these components as element \textit{Interface} ($86\%$, $93\%$, $86\%$, and $79\%$, respectively). In contrast, only $8\%$ of \GDFD{} modelled JTAG, Bluetooth, and Wi-Fi and $25\%$ modelled the USB interface. Since the majority of \GCPDFD{} modelled all these components, while \GDFD{} did not, this suggests an added value of the two elements \textit{Physical Link} and \textit{Interface}. 

Last, a brief review of the questionnaire. The only item of interest in this study was about the simplicity of creating a \gls{CPDFD}/\gls{DFD}. A total of $61.54\%$ of \GCPDFD{} agreed that creating a \gls{DFD} is easy, while $41.67\%$ agreed of \GDFD. This seems to indicate that the two new elements of a \gls{CPDFD} have not increased the difficulty of a \gls{DFD}, but actually made it easier. However, this difference is not significant (\MWUT, $U = 13.5$, $p = .911$, $p > \alpha$). Therefore, the difficulty of creating a \gls{CPDFD} and \gls{DFD} can be considered as equal.

\subsection{Survey and Interviews}
\label{ssec:Survey-Res}
This section presents the results of the survey and interviews with device manufacturers and consulting companies. The precise duration for conducting the study was not measured. Participants indicated that they invested from two hours up to two days. Most of them applied the \gls{TTM} methodology on their own devices. A few also carried out the analysis for other devices, e.g., power inverter or smart home. Some people used the results from previous assessments for a side-by-side comparison.

\paragraph{Questionnaire}
All participants agreed that it is worth to create and analyse a hardware model ($13$ strongly agreed, $2$ agreed) and no one disagreed that reusing the elements of a \gls{DFD} is the right technique for this ($4$ strongly agreed, $5$ agreed, $3$ indecisive). It was also asked about the added value of the two new elements, resulting in exactly the same outcome. The majority of $58.3\%$ strongly agreed that both elements \textit{Physical Link} and \textit{Interface} provided added value ($17\%$ agreed, $8\%$ each indecisive, disagreed, and strongly disagreed). They also agreed ($72.7\%$) that the elements contribute to a more accurate modelling of \gls{IoT} devices. Additional feedback was provided noting that the two elements facilitate modelling, but that the interface needs to be more integrated. It was also remarked that the meaning of both elements is similar. The results suggest that the hardware model, the \textit{Physical Link}, and the \textit{Interface} improve the threat identification. 

\paragraph{Interviews}
This section summarises the findings of the interview analysis. Statements by individual persons are anonymised. If necessary, the person is described by categories introduced in \autoref{ssec:Survey-Meth}. In these cases, the role of the persons and their company type are indicated in parentheses, together with their participant number, e.g., (P\sub{1}, security expert, manufacturer).

When participants were asked about the relevance of a hardware model, all agreed that this is an important part. Several participants stated that the hardware model was the best part of the \gls{TTM} methodology. Hardware attacks are not considered sufficiently and ``without them one cannot find the crux of the matter'' (P\sub{8}, security expert, manufacturer). One participant had tried to represent hardware security in a model in the past using the system modelling language. However, he said that a structured threat analysis was not possible. In general, it was seen as positive that one can quickly see how the device is constructed and which interfaces and components, such as \glspl{TPM}, it uses.

The extension of the \gls{DFD} notation with the two elements \textit{Physical Link} and \textit{Interface} was seen positive, but some participants were not fully convinced. One participant found the extension very good, as she ``struggled in the past to map certain hardware components in a well-distinguishable yet abstract way'' (P\sub{13}, security expert, service provider). The \textit{Physical Link} was found to be good, as data is generated there and actuators can influence the environment. ``Even if a sensor is analogue, that does not mean you don‘t have to think about it'' (P\sub{10}, security expert, service provider). More feedback was given for the \textit{Interface}. It is useful and important, as ``debug interfaces are often overlooked'' (P\sub{9}, security expert, service provider). In a certain way, one is also forced to think about interfaces. However, a few participants noted that they did not fully understand the difference between both elements. The elements could also make the models more abstract and thus more vague in certain cases. One participant noted that the new elements are useful; their use is left to the user.

The feedback from the participants clearly shows that the hardware model brings improvements and is important for the threat analysis. Overall, the two new elements were also seen positively and their enrichment appreciated. 

\section{Discussion}
\label{sec:Discussion}

\subsection{Hardware Model}
In the experimental study, the median of group \GCPDFD{} identified $29.5$ attack scenarios in the hardware model ($59.1\%$ of all), \GDFD{} found at least $11.5$ scenarios ($42.2\%$). The efficiency was also higher for the hardware model ($1$~minute per scenario) than for the average ($2$ and $4$ minutes per scenario for \GCPDFD{} and \GDFD{} respectively). In the survey, all participants, security experts and engineers, agreed that it is worth to create and analyse a hardware model (see \autoref{ssec:Survey-Res}). Furthermore, they reaffirmed the importance of the hardware model in the interviews. One of the advantages they mentioned was the quick overview of all components and interfaces. In conclusion, the quantitative and qualitative results speak in favour of a hardware model. This suggests that using a hardware model may improve threat identification.

\subsection{Physical Link and Interface}
The two new elements \textit{Physical Link} and \textit{Interface} were evaluated separately, but the jointly collected results differ only slightly. Their results are therefore summarised together first and then briefly discussed individually.

In the experimental study, the two groups \GCPDFD{} and \GDFD{} had the single difference that the former had the two new elements available in the diagram editor. However, \GCPDFD{} found significantly more scenarios than \GDFD{} ($49.9$ and $27.3$ on average respectively) (see \autoref{ssec:ExperimentalStudy-Res}). The assignment of the scenarios to elements showed that \GCPDFD{} also discovered $27.9$ scenarios for the standard elements. The additional scenarios were initiated by the \textit{Physical Link} ($10.9$) and the \textit{Interface} ($11.2$). There was also a difference regarding privacy-related threats. Group \GCPDFD{} found on average $5.0$ scenarios while the \GDFD{} found less than $1$ scenario. The difference is not statistically significant, but can be explained by the fact that \GCPDFD{} found $3.6$ scenarios with the \textit{Physical Link} and $1.2$ scenarios with the \textit{Interface}. This effect could also be seen in the hardware model. The median found $29.5$ and $11.5$ scenarios for \GCPDFD{} and \GDFD{}, respectively. More than half of the scenarios of the former were based on the \textit{Physical Link} ($6$) and \textit{Interface} ($10.5$). It was also analysed how participants modelled specific components. For example, almost all of group \GCPDFD{} modelled the microphone as \textit{Physical Link}. In contrast, only $42\%$ of \GDFD{} modelled it. They were indecisive which element to use, as $3$ participants modelled it as \textit{Process} and $2$ as \textit{External Entity}. The interfaces such as JTAG and Bluetooth were modelled by about $90\%$ of \GCPDFD{} while these were modelled by only one third or less of \GDFD. This means that the existence of the two new elements leads to such components being modelled at all. The two elements do not make the technique more difficult, as the questionnaire showed. In the survey, the majority of $75\%$ agreed that each \textit{Physical Link} and \textit{Interface} provide added value and $73\%$ agreed that they contribute to more accurate modelling of \gls{IoT} devices (see \autoref{ssec:Survey-Res}). In the interviews, the elements were generally seen positive. One security expert confirmed the problems with modelling certain hardware components without the new elements. The \textit{Physical Link} was said to be good as data is generated there and actuators can influence the environment and the \textit{Interface} was conceived to be important as debug interfaces are often overlooked, among others. However, it was noted a few times that the differentiation between both elements is not clear and that they could be summarised in one element. In summary, there are many arguments in favour of introducing the two new elements. The \textit{Physical Link} leads to several additional scenarios, as seen in the experimental study, although not as many as for the \textit{Interface}. The element can be used to represent sensors, which in turn have mainly contributed to the identification of threats against privacy. This added value was also endorsed by the survey participants. The results thus suggest that the \textit{Physical Link} improves the threat identification. The situation is similar for the \textit{Interface}. Quantitatively, even more scenarios could be identified with this element. The importance of interfaces was highlighted by students as well as security experts and engineers. Therefore, the results suggest the \textit{Interface} improves the threat identification as well.

\subsection{Limitations}
Some limitations must be taken into account in the reported results. The \gls{HWD} produced dozens of attack scenarios, but several of them may also be found in \glspl{DFD}, creating overlaps. Consequently, the number of additional scenarios contributed by the \gls{HWD} is lower than the raw count suggests. The quantitative effects of the two new elements must also be considered with caution. In the experimental study, participants of \GCPDFD{} had predefined stencils for the \textit{Physical Link}, e.g., microphone and loudspeaker, and for the \textit{Interface}, e.g., JTAG and Bluetooth. The list of dozens of stencils had to be searched through first, but these increase the likelihood that such components will be modelled. The modelling then automatically led to further scenarios. For a more detailed examination of the effects of the two elements, further groups would have been necessary, which would not have had predefined stencils available.

\section{Conclusions and Future Work}
\label{sec:Conclusions}
The many security incidents in the \gls{IoT} highlight the ongoing need for improved security. Although numerous researchers address \gls{IoT} security, systematic approaches are often lacking. This work introduces a modelling technique tailored to \gls{IoT} devices to simplify modelling and enhance threat identification, helping manufacturers address weaknesses early in the product development cycle and support security by design. The proposed \gls{CPDFD} technique considers the special circumstances of \gls{IoT} devices. Due to their use in critical applications as well as the high accessibility, the creation of a hardware model was proposed. In addition, due to the interaction with the environment and the high connectivity of \gls{IoT} devices, the two new elements \textit{Physical Link} and \textit{Interface} were introduced. Across two studies, an experimental quantitative comparison was conducted and the perspectives of security experts and engineers were collected. In summary, all three changes of the \gls{CPDFD} technique indicate improvements compared to the standard \gls{DFD}. The technique enables more detailed modelling of \gls{IoT} devices, which in turn leads to more identified attack scenarios, as the results demonstrated. This suggests that \glspl{CPDFD} improve the identification of threats. In the future, the better integration of \glspl{HWD} and \glspl{DFD} will be addressed to exploit synergy effects and improve the identification of attack scenarios and countermeasures.

\balance
\printbibliography

\end{document}